\documentclass[%
 11pt,
 reprint,
superscriptaddress,
preprintnumbers,
 amsmath,amssymb,
]{revtex4-2}

\usepackage{graphicx}
\usepackage{dcolumn}
\usepackage{bm}
\usepackage{subcaption} 
\usepackage{booktabs} 
\usepackage{multirow} 
\usepackage{color}
\usepackage[newcommands]{ragged2e}
\usepackage{accents}

\newcommand{\sta}{{~}\,{{{\bullet}}}\,{~}}

\newcommand{\stb}{{~}\,{\underline{{\bullet}}}\,{~}}

\newcommand{\stc}{\underline{~}\,{{\bullet}}\,\underline{~}}

\newcommand{\std}{\underline{~}\,{\underline{{\bullet}}}\,\underline{~}}

\newcommand{\ste}{\underaccent{\sim}{~}\,{ \underline{{\bullet}}}\;\underaccent{\sim}{~}}

\newcommand{\stf}{{~}\,{{\overline{\bullet}}}\,{~}}

\newcommand{\stg}{{~}\,{\underline{\overline{\bullet}}}\,{~}}

\newcommand{\sth}{\underline{~}\,{\overline{\bullet}}\,\underline{~}}

\newcommand{\sti}{\underline{~}\,{\underline{\overline{\bullet}}}\,\underline{~}}

\newcommand{\stj}{\underaccent{\sim}{~}\,\accentset{\sim}{ \underline{\overline{\bullet}}}\;\underaccent{\sim}{~}}

\newcommand{\stk}{\underline{~}\,\underaccent{\sim}{\overline{\bullet}}\,\underline{~}}

\begin{document}


\title{Multiple Equilibria Enables Tunable Wetting of Droplets on Patterned Liquid Surfaces}

\author{Xitong Zhang}
\thanks{These authors contribute equally.}
\affiliation{Institute for Multiscale Thermofluids, School of Engineering, The University of Edinburgh, King's Buildings, Mayfield Road, Edinburgh EH9 3FD, U.K.}
\affiliation{Department of Physics, Durham University, South Road, Durham DH1 3LE, U.K.}
\author{Hongyu Zhao}
\thanks{These authors contribute equally.}
\affiliation{Institute for Multiscale Thermofluids, School of Engineering, The University of Edinburgh, King's Buildings, Mayfield Road, Edinburgh EH9 3FD, U.K.}
\author{Jack R. Panter}
\affiliation{School of Engineering, University of East Anglia, Norwich, Norfolk NR4 7TJ, UK.}
\author{Glen McHale}
\email{glen.mchale@ed.ac.uk}
\affiliation{Institute for Multiscale Thermofluids, School of Engineering, The University of Edinburgh, King's Buildings, Mayfield Road, Edinburgh EH9 3FD, U.K.}
\author{Gary G. Wells}
\email{gary.wells@ed.ac.uk}
\affiliation{Institute for Multiscale Thermofluids, School of Engineering, The University of Edinburgh, King's Buildings, Mayfield Road, Edinburgh EH9 3FD, U.K.}
\author{Rodrigo Ledesma-Aguilar}
\email{rodrigo.ledesma@ed.ac.uk}
\affiliation{Institute for Multiscale Thermofluids, School of Engineering, The University of Edinburgh, King's Buildings, Mayfield Road, Edinburgh EH9 3FD, U.K.}
\author{Halim Kusumaatmaja}
\email{halim.kusumaatmaja@ed.ac.uk}
\affiliation{Institute for Multiscale Thermofluids, School of Engineering, The University of Edinburgh, King's Buildings, Mayfield Road, Edinburgh EH9 3FD, U.K.}
\affiliation{Department of Physics, Durham University, South Road, Durham DH1 3LE, U.K.}

\date{\today}

\begin{abstract}
{Patterning solid surfaces with varying wettability is important to manage droplets
in microfluidics, heat transfer and printing. Solid surface roughness poses fundamental limitations including
contact-line pinning and solid friction. Here, we report an experimental strategy and theoretical
design principles for Patterned Liquid Surfaces (PaLS) that combines the controlled
wettability from patterning with the ultra-smoothness of a lubricant-infused surface. In contrast to
a solid, on PaLS a droplet can be in 10 different wetting states. This richness arises from
the adaptation of the liquid lubricants, and can be harnessed to control the apparent contact angle
of the droplet over the full range of wettability whilst removing contact-line pinning effects induced
by the solid surface. In the limit of thin liquid films, we derive surface-averaged
laws for the apparent contact angle for each wetting state, which capture both experimental and
simulation data. Our results provide a distinct approach to surface patterning that exploits the interaction of fluids with lubricant-impregnated surfaces. }
 
\end{abstract}

\maketitle

\section{Introduction}
\label{sec:level1}
The ability to engineer the interaction of liquid droplets with solid substrates is important in many applications, including microfluidics~\cite{malinowski2020advances, zeng2024MaterialsTodayPhysics}
, heat transfer~\cite{
tokunaga2020enhancement}, inkjet printing~\cite{lohse2022fundamental} and surface cleaning~\cite{mazaltarim2021dynamic}.
%
%
Patterning a solid substrate, either chemically or topographically, is perhaps the most popular approach to controlling a droplet's shape, position, and motion. 
Indeed, patterning occurs in many biological systems where droplets come into contact with a solid, and such systems have inspired the design of solid surfaces for specific applications.
For example, Namib Desert beetles employ patches of hydrophilic and hydrophobic regions of different curvatures that facilitate water harvesting~\cite{parker2001water}
and Araucaria leaves have asymmetric structures that result in unidirectional liquid motion~\cite{feng2021three}. 
%
%
However, solid surface patterns have a fundamental limitation. Solid surfaces are rough, leading to difficulties in controlling contact-line pinning and strong solid friction forces {\cite{quere2008wetting,bonn2009wetting}}.

Porous or textured solid substrates infused with a liquid lubricant layer, often called Slippery Liquid-Infused Porous Surfaces (SLIPS)~\cite{wong2011bioinspired}, slippery pre-suffused surfaces~\cite{lafuma_slippery_2011}, or Lubricant-Impregnated Surfaces (LIS)~\cite{smith2013droplet}, 
offer the distinct advantage of reducing contact between the droplet and the solid substrate, decreasing contact-line pinning, and enhancing droplet mobility. Such lubricant-infused surfaces also exhibit self-healing, pressure stability and optical transparency, thus making them attractive in applications such as surface cleaning, anti-icing \cite{kim2012liquid, wilson2013inhibition}, fog collection \cite{qi2022ultraslippery}, and heat transfer \cite{anand2012enhanced, xiao2013immersion}. 

{\color{black} {\color{black}The concept of PaLS (Patterned Liquid Surfaces) -- solid surfaces impregnated by small but distinct domains of two immiscible liquid
lubricants -- brings together two powerful strategies in surface engineering, combining the benefits of controlled wettability from patterning with the ultra-smoothness of lubricant-infused surfaces.}
At the same time, PaLS bring to the fore several new fundamental questions in wetting phenomena and multiphase fluid mechanics. On an equivalent patterned solid surface, the configuration of a droplet is determined by a surface average of the individual chemical components of the pattern. In contrast, the adaptability of the liquid lubricants can potentially give rise to new droplet states. On a lubricant-infused surface, a droplet is known to adopt a well-defined apparent contact angle~\cite{mchale2019apparent,Semprebon2017SoftMatter}, which is determined by the interfacial energies between the solid, the lubricant, the droplet, and the surrounding gas phase~\cite{smith2013droplet}. While some variation of the contact angle can be achieved by choosing different solid/lubricant/droplet combinations~\cite{mchale2019apparent,mchale2022liquid}, or by introducing a topographical micro-pattern in the solid~\cite{sadullah2020bidirectional,launay2020self}, the range is limited and the resulting surfaces can suffer from contact-line pinning because of contact between the droplet and the solid. Can we overcome these limitations using PaLS?

%
%
%
%
%
%

\begin{figure*}[t!]
    \centering
    \includegraphics[width=\textwidth]{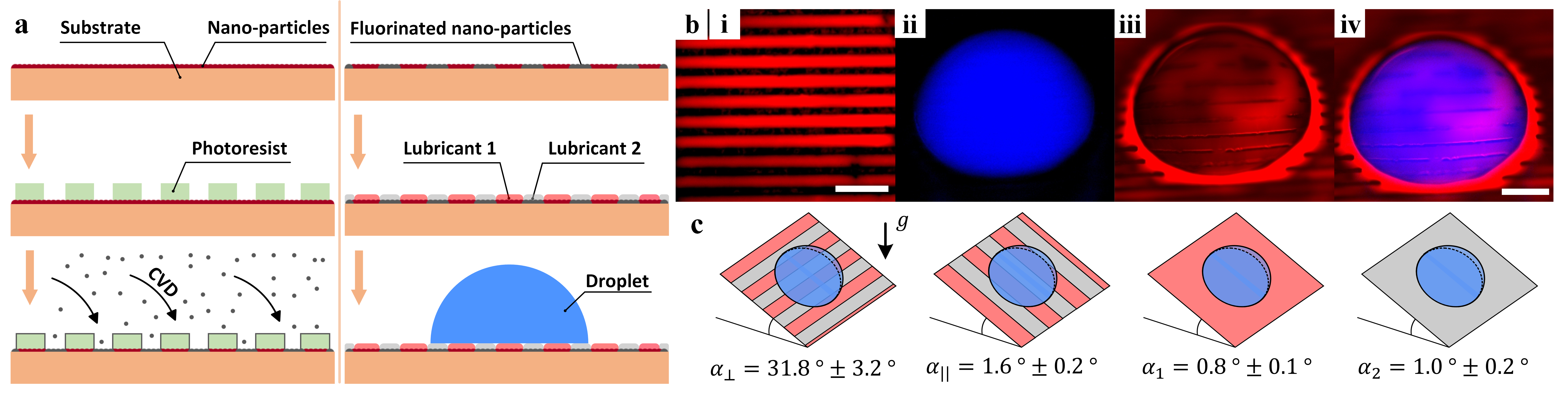}
    \captionsetup{justification=Justified,width=\textwidth} 
    \caption{{\bf Creating Patterned Liquid Surfaces (PaLS).} 
    {\bf a} Surface fabrication procedure. 
    A glass substrate is coated with a thin nano-particle layer and covered with a potoresist to create an exposed surface pattern. 
    Chemical vapour deposition (CVD) is used to fluorinate the exposed areas, creating a chemical pattern in the nano-particle layer upon removal of the photoresist. 
    Immiscible liquid lubricants are applied to create a PaLS able to support a droplet of a third liquid in air or a gas bubble surrounded by another liquid. 
    {\bf b} Fluorescence microscopy images of a PaLS formed by olive oil/Krytox $50\mu{\rm m}$-wide stripes (i). Olive oil is dyed red, Krytox appears black and a water droplet is dyed blue. Composite channel images show cloaking of the water droplet by olive oil (ii)-(iv).  The scale bars are 500~$\mu{\rm m}$.
    {\bf c} Sliding angles for a 10-$\mu{\rm L}$ water droplet in the perpendicular and parallel directions to the stripes, $\alpha_{\perp}$ and $\alpha_{\parallel}$, and sliding angles on surfaces coated by olive oil and Krytox only, $\alpha_{1}$ and $\alpha_2$, respectively. 
    }
    \label{fig:Fig1}
\end{figure*}

%
%
To address these open challenges, we study 
liquid droplets and gas bubbles on PaLS.
%
%
Our experimental method to create the surfaces departs from previous approaches~\cite{vogel2013transparency,ueda2013micropatterning,paulssen2019droplet,pelizzari2023droplet} in achieving the coating of a solid substrate with micro-scale domains of two immiscible lubricants without the need of solid barriers. 
This creates a small-scale liquid pattern free of solid pinning whose wettability can be adjusted by selecting its composition and geometry.
We show that, on PaLS, a liquid droplet or an immersed gas bubble can take 10 distinct wetting states, including complex configurations that depend on whether the lubricants cloak each other, and whether the droplet's surface remains exposed or is cloaked by one or both lubricants.
%
Using a combination of experiments, numerical simulations and analytical modelling, we demonstrate how choosing the lubricants according to their physical properties can be used to select a specific wetting state, whilst changing the surface fraction of each lubricant gives us control over the apparent contact angle.
Interestingly, the classical Cassie-Baxter law~\cite{cassie_wettability_1944} fails to capture the apparent contact angle on PaLS. To remedy this failure, we derive a new surface averaged law for the apparent contact angle that accounts for the complex lubricant cloaking behaviors on PaLS.
}

\section{Results}
\subsection{Fabrication and Visualization of Patterned Liquid Surfaces}
We developed a new method to create PaLS by coating a solid substrate with immiscible lubricant domains arranged in a prescribed geometry (see Fig.~\ref{fig:Fig1}a).  
Full details of the surface preparation are provided in the Methods section. 
Briefly, pre-cleaned glass slides 
were coated with Glaco$^{\rm TM}$, a silica-based superhydrophobic nano-particle coating, 
to create a thin porous layer. 
A sacrificial layer of photoresist 
was deposited on the samples using spin coating. 
The photoresist was etched using photolithograpy to create the desired pattern.  
The surfaces were fluorosilanized via chemical vapor deposition (CVD) before the sacrificial photoresist was removed. 
The resulting nano-particle layer comprises bare Glaco$^{\rm TM}$ (G) and fluorosilanized (FS) domains.
Glaco$^{\rm TM}$ domains have preferential affinity for alkane-based or siloxane-based lubricants, while fluorosilanized domains have affinity for fluorocarbon lubricants~\cite{mchale2022liquid,pelizzari2023droplet}. 
Sequential dip-coating of the surface with suitable lubricant combinations creates a stable surface with well-defined lubricant domains, which in turn can support a droplet of a third liquid (or, if desired, a gas bubble surrounded by another liquid). 
Our method can be used to create domains of arbitrary geometry at high resolution (up to $\approx 5~{\mu \rm m}$). 

Here, we study the interaction of PaLS with small liquid droplets in air or bubbles in liquids employing a parallel-stripe geometry formed by alternating domains of Krytox, a fluorocarbon oil, 
and either food-grade olive oil or silicone oil 
(see Fig.~\ref{fig:Fig1}b). 
Taking advantage of the high resolution of our surface fabrication method, we vary 
the width of the stripes between approximately $5$ and $450~{\rm \mu m}$, which is small compared to the base diameter of the droplet ($\approx 2~{\rm mm}$). 
The pattern geometry is characterized by the surface fraction of flurosilanized domains, $f\equiv w_{\rm FS}/(w_{\rm FS}+w_{\rm G})$, where $w_{\rm G}$ and $w_{\rm FS}$ are the widths of Glaco$^{\rm TM}$ and fluorosilanized stripes, respectively.

\begin{figure*}
    \centering
    \includegraphics[width=\textwidth]{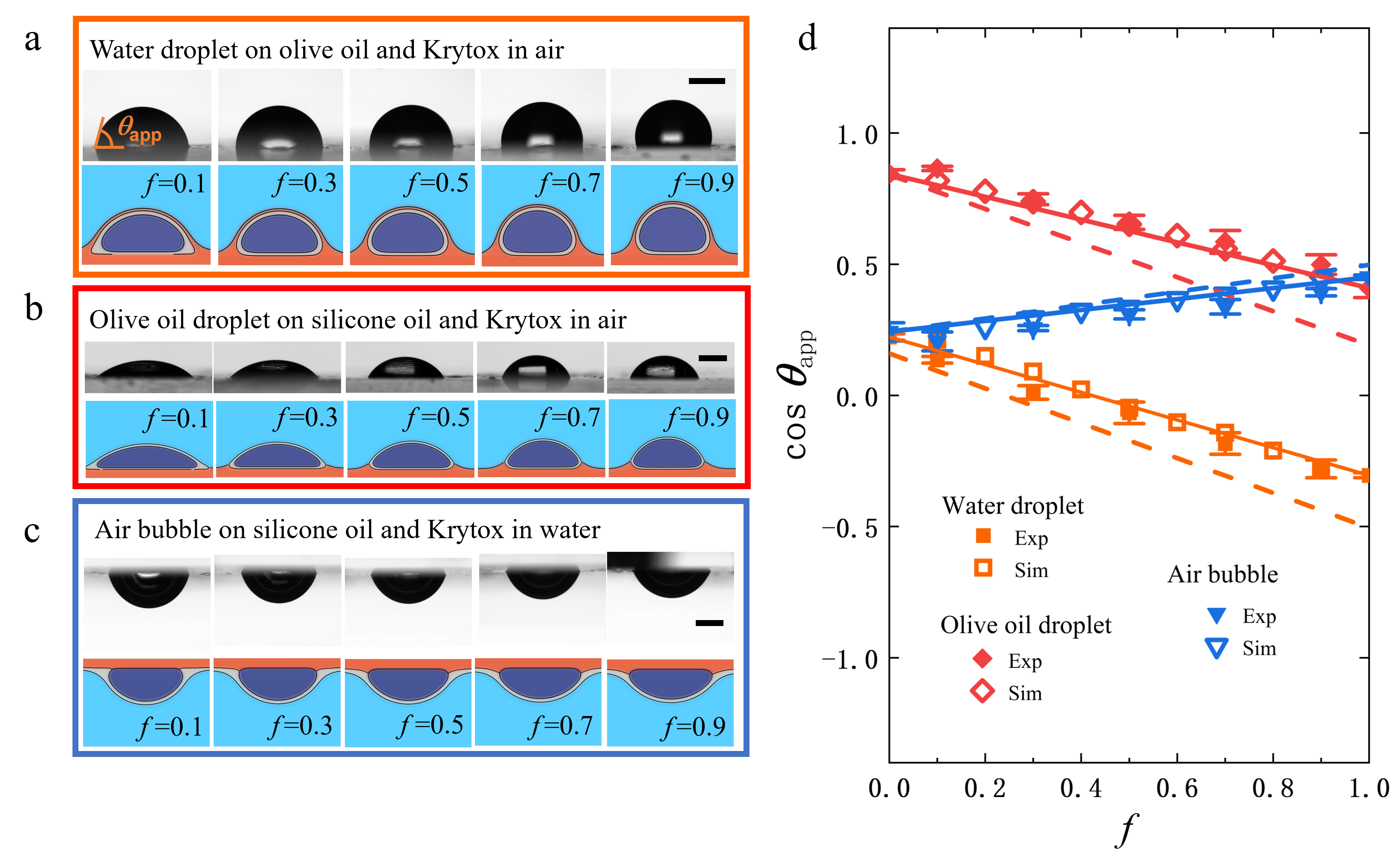}\captionsetup{justification=Justified,width=\textwidth} 
    \caption{{\bf Controlling the wettability of PaLS.} {\bf a-c}~Top rows: Experimental images of droplets and bubbles on patterned liquid surfaces composed of stripes of immiscible lubricants at different stripe fraction, $f$.
    Bottom rows: cross-sections of droplets/bubbles obtained from numerical simulations based on free-energy minimization.
    {\bf a}~Water droplet on a Krytox-olive oil surface surrounded by air. 
    {\bf b}~Olive oil droplet on a Krytox-silicone oil surface surrounded by air.
    {\bf c}~Air bubble on a Krytox-silicone oil surface surrounded by water.
    {\bf d}~Variation of the cosine of the apparent contact angle with surface fraction. The full and empty markers correspond to experimental and simulation results, respectively.
    The dashed lines correspond to the prediction of the classical Cassie-Baxter law that neglects the effect of both lubricants, while the solid lines are from surface average equations that include the effect of both lubricants.
    The scale bars in {\bf a-c} are 1~mm.}
    \label{fig:Fig2}
\end{figure*}


When placed on the surface, a water droplet adopts a stable equilibrium shape which depends on the choice of the lubricants and the surface fraction (Fig.~\ref{fig:Fig1}b). 
Tilting the surface along the direction of the stripes results in very low sliding angles ($\alpha_\parallel\approx 1^\circ$), comparable to the sliding angles measured separately on surfaces coated by either lubricant~(Fig.~\ref{fig:Fig1}c), thus indicating minimal pinning from the underlying solid. 
In sharp contrast, tilting the surface in the transverse direction to the stripes results in much higher sliding angles~(e.g., $\alpha_\perp \approx 30^\circ$ for the combination reported in the figure), showing that the lubricant domains introduce a strong anisotropic capillary interaction with the droplet. Details of the sliding angle measurement protocol is provided in the Supplementary Materials. 

Fig.~\ref{fig:Fig1}b(ii-iv) illustrates the complexity of the interfacial configuration of a droplet on a patterned liquid surface. 
The droplet (dyed in blue) spreads anisotropically due to the stripe pattern formed by olive oil (dyed in red) and Krytox (undyed, shown in black). 
Olive oil collects to form a rim around the droplet and cloaks the droplet's surface. 
While providing useful information, direct visualization using dyes can be impractical for several reasons. 
First, droplet cloaking, as is the case shown in the figure, makes it difficult to clearly determine the lubricant configuration underneath the droplet. 
Second, one needs a third dye to visualize the second lubricant (Krytox). 
More generally, the dyes must be chosen carefully depending on the liquids employed, such that each dye is only miscible in one of the phases and they do not significantly alter the interfacial tensions. 
The latter can lead to undesirable changes in the droplet wetting state.

\begin{figure*}
    \centering
    \captionsetup{justification=Justified,width=\textwidth}
    \includegraphics[width=\textwidth]{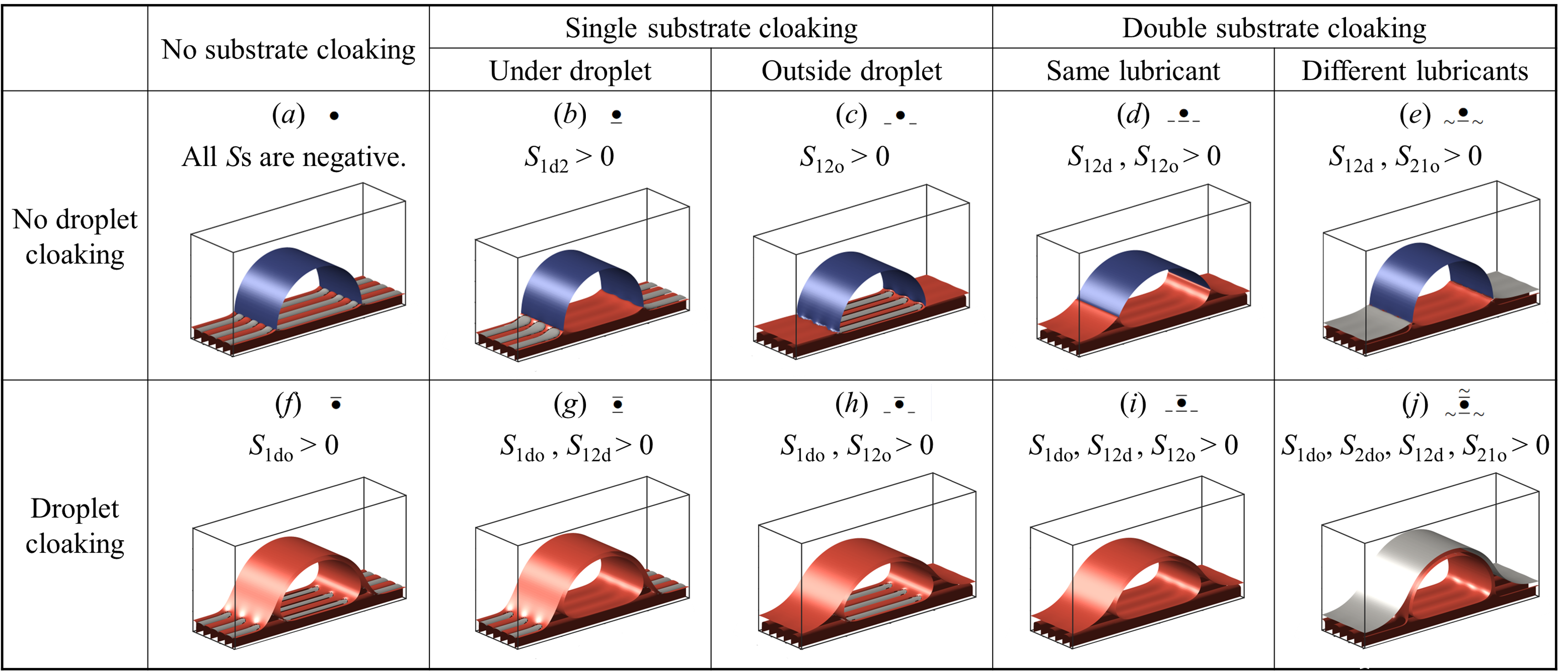}
    \caption{ {\bf Equilibrium states of droplets on PaLS.} 
    The ten possible equilibrium states can be classified into states where no droplet cloaking occurs (states ($a$)-($e$)), and where it does (states ($f$)-($j$)). 
    Substrate cloaking (where one lubricant cloaks the other) can occur underneath the droplet (($b$) and ($g$)), 
    outside of it (($c$) and ($h$)), or in both locations (($d$) and ($i$)).
    Cloaking by different lubricants underneath and outside the droplet can occur in two cases: ($e$) and ($j$). 
    Double cloaking occurs in state ($j$).
    The images correspond to equilibrium simulation configurations obtained from energy minimization of a 4-phase fluid model with $f=0.5$. 
    }
    
    \label{fig:wetting state} 
\end{figure*}   

These observations open several questions. 
What determines the shape of the droplet? How are the different phases in the system configured in equilibrium? 
To address these questions, we varied the surface fraction of flurosilanized stripes, $f$, and studied the shape of the droplet by measuring the apparent contact angle $\theta_{\rm app}$ (Fig.~\ref{fig:Fig2}a). 
In the experiments, the lubricants are in thin films and so we define the
apparent contact angle by fitting the interface shape to
a circular profile whose slope is extrapolated at the level
of the flat surface (see Supplementary Materials).
As shown in the figure, increasing the surface fraction has the effect of increasing the apparent angle. 
We repeated the same experiment for different combinations of the fluids, including situations where the droplet is replaced by an immersed gas bubble.
Examples of resulting droplet/bubble configurations are shown in Figs.~\ref{fig:Fig2}b-c (a full list of the different cases considered in this work is reported in Table~\ref{tab:spreading parameter}).  
In all cases, the apparent contact angle varies monotonically as the surface fraction changes, suggesting that it is determined by the average effect of the wettability of each lubricant component.

From classical wetting theory on solid substrates, one might expect that the apparent contact angle is governed by Cassie-Baxter's law~\cite{cassie_wettability_1944}, 
\begin{equation}
\label{eq:CassieBaxter}
\cos \theta_{\rm CB}=\cos \theta_{\rm 1} + f(\cos\theta_{\rm 2}-\cos\theta_1),
\end{equation}
where the predicted apparent contact angle, $\theta_{\rm CB}$, results from the spatial average of the contact angles of a droplet observed separately on each lubricant, $\theta_1$ and $\theta_2$, corresponding to $f = 0$ and $f = 1$.  
These angles can be measured independently.
However, using such ``bare" values for $\theta_1$ and $\theta_2$ in Eq.~\eqref{eq:CassieBaxter} fails to predict the observed apparent contact angle, as shown by the dashed lines in Fig.~\ref{fig:Fig2}b. 
Instead, we find that our data can be captured only if we employ $\theta_1$ and $\theta_2$ that are measured when both lubricants are present (solid lines in the figure), indicating that both lubricants influence the droplet configuration over the full range of surface fractions, even at trace amount when $f \rightarrow 0$ and $f \rightarrow 1$.   
%

%
%
%
%

\subsection{Droplet Wetting States}

To better understand the lubricant configurations and droplet wetting states, we now develop a theoretical model based on the spreading parameters of the liquids and turn to computer simulations using a phase field method to systematically verify the model predictions and illustrate the possible equilibrium states. 



A useful quantity to predict the configuration of the four fluid phases in equilbirium is the spreading parameter. 
Given three different fluid phases ($i$, $j$, and $k$), we define the spreading parameter of fluid $i$ at the interface between fluids $j$ and $k$ as $S_{ijk}=\gamma_{jk}-(\gamma_{ij}+\gamma_{ik})$. 
Here, $\gamma_{ij}$ denotes the interfacial tension between fluids $i$ and $j$. 
A positive value of $S_{ijk}$ indicates that a single interface between phases $j$ and $k$ is more energetically expensive than the two interfaces between phases $i-j$ and $i-k$. 
When this happens, it is preferable for fluid $i$ to completely spread at the interface between phases $j$ and $k$, i.e., we say that fluid $i$ cloaks the $j-k$ interface, or, in other words,  that fluid $i$ cloaks fluid $j$ in the presence of fluid $k$.

Let us now identify the possible equilibrium states arising from the combination of the cloaking configurations between the lubricants and the droplet.
Without loss of generality, we will assume that lubricant 1 (labeled ``1''; colored red in Fig.~\ref{fig:wetting state}) has a larger spreading coefficient than lubricant 2 (``2''; colored gray) at the interface between the droplet (``d'') and outer fluid phase (``o'').
In this section, we focus on equilibrium states where the droplet neither completely spreads on nor completely detaches from either lubricant, i.e.,  $S_{{\rm d}1{\rm o}}<0$, $S_{{\rm d}2{\rm o}}<0$, $S_{{\rm o}1{\rm d}}<0$
and $S_{{\rm o}2{\rm d}}<0$ (we shall later discuss situations where these constraints are relaxed).
Importantly, within these constraints, the lubricants can still cloak the droplet and/or each other. Combinations of the spreading coefficients define 10 unique wetting states, shown in Fig.~\ref{fig:wetting state}.
In terms of classification hierarchy, we first group these states depending on whether cloaking of the droplet occurs.
States ($a$)-($e$) correspond to uncloaked droplet states ($S_{1{\rm do}}<0$) while states ($f$)-($j$) correspond to cloaked droplet states ($S_{1{\rm do}}>0$). 
Within this first classification, one also needs to consider whether, and where, lubricant-lubricant cloaking occurs.
For states ($a$) and ($f$) no lubricant-lubricant cloaking occurs. 
In states ($b$) and ($g$), lubricant cloaking occurs underneath the droplet; whereas in states ($c$) and ($h$), it occurs outside the droplet. 
In states ($d$) and ($i$), cloaking by the same lubricant occurs both underneath and outside the droplet. 
%
%
However, in states ($e$) and ($j$), different lubricants cloak underneath and outside the droplet: 
lubricant 1 cloaks lubricant 2 underneath the droplet, and vice versa outside the droplet. 
In addition, state ($j$) shows double cloaking~\cite{pelizzari2023droplet}, so that a layer of lubricant 1 cloaks the droplet, and this is in turn cloaked by lubricant 2.

It is useful to identify the equilibrium states using vignettes, where ``$\bullet$'' represents the droplet, and ``$-$'' and ``$\sim$'' are used to indicate cloaking by lubricants 1 and 2 at specific positions. 
For example, state $(h)$, where cloaking by lubricant 1 occurs on the droplet and outside, but not underneath, is symbolized as $\sth$.

We found that no additional wetting states to those reported in Fig.~\ref{fig:wetting state} are possible without raising inconsistencies in the signs of the spreading coefficients. 
For example, consider a hypothetical state akin to state $\sth$ but with lubricant 2 (gray) cloaking lubricant 1 (red) underneath the droplet, i.e., $\stk$. 
Since lubricant 2 does not cloak the droplet, we have $S_{\rm 2do}<0$, or  
$\gamma_{\rm do} < \gamma_{\rm d2} + \gamma_{\rm 2o}$. 
Furthermore, since lubricant 1 cloaks both the droplet and lubricant 2 outside the droplet, we have $\gamma_{\rm do} > \gamma_{\rm d1} + \gamma_{\rm 1o}$ and $\gamma_{\rm 2o} > \gamma_{\rm 21} + \gamma_{\rm 1o}$. 
Combining these inequalities yields the condition $\gamma_{\rm d1} < \gamma_{\rm 2d} + \gamma_{\rm 21}$, or $S_{\rm 21d}<0$, which implies that lubricant 2 cannot cloak lubricant 1 underneath the droplet. 

To analyze the stability of the wetting states, we carried out numerical simulations of a phase-field model that resolves the interaction between the 4 fluid phases.
Stable states were found by minimizing the free energy
\begin{equation}
    E= E_{\text{bulk}}+ E_{\text{surf}}+ E_{\text{conf}} + E_{\text{ens}},
    \label{total free energy}
\end{equation}
where $E_{\text{bulk}}$ and $E_{\text{surf}}$ correspond to the bulk and interfacial energy contributions, which allow the coexistence of the 4 immiscible fluid phases; 
$E_{\text{conf}}$ represents the energy potential of the pattern, which allows us to confine the lubricants in separate domains; and $E_{\text{ens}}$ takes into account the constant-volume ensemble for the droplet and constant-pressure ensemble for the lubricants. 
%
%
The minimization of $E$ was carried out using the L-BFGS algorithm \cite{liu_limited_1989, nocedal_updating_1980}, which is routinely used in optimization problems with a large number of degrees of freedom \cite{panter_harnessing_2019, li_reconfiguration_2021}. 
%
Further details on the free energy model are provided in the Methods section.

Snapshots of the droplet configurations obtained from the numerical simulations are shown Fig.~{\ref{fig:wetting state}}.
We model droplets of cylindrical cross-section oriented parallel to the direction of the stripes~\cite{mognetti2010drop}.
While the model can be used to model full 3D geometries (see Supplementary Materials for examples), this cylindrical simulation geometry is less expensive computationally. It allows us to capture the effect of the lubricant pattern on the droplet shape and explore a broad range of parameters.
%
%
The simulations capture the 10 wetting states expected from the theoretical prediction and confirm that no further configurations involving the 4 fluid phases are possible.
%
%
The simulations also reveal the presence of lubricant wetting ridges surrounding the droplets, 
similar to the ridges observed for droplets on surfaces coated with a single lubricant~\cite{schellenberger2015direct}. 
However, the interaction with two lubricants introduces undulations of the ridge along the inner and outer edges of the droplet.
For example, this effect is particularly noticeable for state $\stf$ in Fig.~{\ref{fig:wetting state}}, where one lubricant cloaks the droplet but not the other, either outside or underneath. 

Fig.~\ref{fig:Fig2} also shows the simulation results for the same droplet (or bubble) and lubricant liquids used in the experiments. Fig.~\ref{fig:Fig2}a-c show the interfacial profiles obtained from the simulations, while the apparent contact angle, and its dependence on the surface pattern fraction, is provided in Fig.~\ref{fig:Fig2}d. The excellent agreement with experimental data underscores the validity of the simulation approach to capture the effect of PaLS on the droplet configuration. At the same time, the simulations provide detailed information on lubricant cloaking and the corresponding interface shapes, which is difficult to obtain in experiments.
%
%

\subsection{Apparent Contact Angle}

\begin{table*}[t] 
 \begin{ruledtabular}
 \renewcommand{\arraystretch}{1.5}
    \begin{tabular}{cccccc}
    \multicolumn{3}{c}{No cloaking}& \multicolumn{3}{c}{Cloaking}\\
    \cmidrule(lr){1-3} 
    \cmidrule(lr){4-6} 
    \multicolumn{1}{c}{State}& \multicolumn{1}{c}{ID} &\multicolumn{1}{c}{$\cos\theta_{\rm app}$} & \multicolumn{1}{c}{State}& \multicolumn{1}{c}{ID} &\multicolumn{1}{c}{$\cos\theta_{\rm app}$}\\
    \cmidrule(lr){1-6} 
    $(a)$ & $\sta$ &$[f(\gamma_{\rm o1}-\gamma_{\rm d1})+(1-f)(\gamma_{\rm o2}-\gamma_{\rm d2})]/\gamma_{\rm do}$   & $(f)$ & $\stf$ & $[f(\gamma_{\rm 1o}-\gamma_{\rm d1})+(1-f)(\gamma_{\rm o2}-\gamma_{\rm d2})]/(\gamma_{\rm d1}+\gamma_{\rm o1})$\\
    $(b)$ & $\stb$ & $[f\gamma_{\rm o1}-\gamma_{\rm d1}+(1-f)(\gamma_{\rm o2}-\gamma_{12})]/\gamma_{\rm do}$   & $(g)$ & $\stg$ &
    $[f\gamma_{\rm o1}-\gamma_{\rm d1}+(1-f)(\gamma_{\rm 2o}-\gamma_{12})]/(\gamma_{\rm d1}+\gamma_{\rm o1})$\\
    $(c)$ & $\stc$ &$[\gamma_{\rm o1}-f\gamma_{\rm d1}+(1-f)(\gamma_{12}-\gamma_{\rm d2})]/\gamma_{\rm do}$   & $(h)$ &  $\sth$ &$[\gamma_{\rm o1}-f\gamma_{\rm d1}+(1-f)(\gamma_{12}-\gamma_{\rm d2})]/(\gamma_{\rm d1}+\gamma_{\rm 1o})$\\
    $(d)$ & $\std$ &$(\gamma_{\rm o1}-\gamma_{\rm d1})/\gamma_{\rm do}$   & $(i)$ & $\sti$ & $(\gamma_{\rm 1o}-\gamma_{\rm d1})/(\gamma_{\rm d1}+\gamma_{\rm o1})$\\
    $(e)$ & $\ste$ & $[\gamma_{\rm o2}+(2f-1)\gamma_{12}-\gamma_{\rm d1}]/\gamma_{\rm do}$   & $(j)$ & $\stj$ & $[\gamma_{\rm o2}-\gamma_{\rm d1}+(2f-1)\gamma_{12}]/(\gamma_{\rm d1}+\gamma_{12}+\gamma_{\rm o2})$\\
    \end{tabular}
    \end{ruledtabular}
    \captionsetup{justification=Justified,width=\textwidth} 
    \caption{\label{tab:table equation}
    {\bf Apparent contact angle laws for PaLS.} 
    The subscripts in the expressions stand for d = droplet, o = outer fluid, 1 = lubricant 1, and 2 = lubricant 2.}
 \end{table*}

%

Our results show that the wettability of patterned liquid surfaces, measured through the apparent contact angle, does not obey the Cassie-Baxter law, but follows a surface averaged law that strongly depends on the specific wetting state adopted by the droplet. 
In this section, we derive a prediction of the apparent contact angle that accounts for the configuration of the full 4-phase system. 
We focus on the regime where the wetting ridge is small compared to the droplet.
The experiments show that droplets spread anisotropically on the striped surfaces. 
Perpendicular to the stripes, the apparent contact angle varies significantly depending on the location of the edge of the droplet relative to the position of the stripes. 
This is similar to the hysteretic behavior observed for droplets on solid surfaces patterned with chemical stripes and parallel grooves~\cite{Bliznyuk2009PRE,Kusumaatmaja2008Langmuir}. 
Parallel to the stripes, the apparent contact angle converges to a unique value. 
This is due to a stronger ability of the droplet edges to spread along the direction of the stripes (as in the configurations shown in Fig.~\ref{fig:wetting state}).
Close to the droplet edge, the shape of the interface is determined by the balance of interfacial tension forces. 
For a droplet on a lubricant-infused surface, where there is only one lubricant, these forces correspond to the surface tensions of the interfaces between the lubricant and the outer fluid, the lubricant and the droplet, and the surface of the droplet~\cite{mchale2019apparent,Semprebon2017SoftMatter}. 
For PaLS, the corresponding interfaces are between the PaLS and the outer fluid, the PaLS and the droplet, and the droplet's surface.
Each of these interfaces has an effective interfacial tension determined by the cloaking configuration and the surface pattern fraction. 


\begin{table*} [t!]
\begin{ruledtabular}
\begin{tabular}{crrrrrrc}
$\rm d(12)o$ & $S_{\rm 1do}  $& $S_{\rm 2do}$ & $S_{\rm 1d2}$ & $S_{\rm 2d1}$  & $S_{\rm 12o}$ & $S_{\rm 21o}$ & Wetting state\\
\hline
w(ok)a & $+22.5 \pm \ 0.5$ & $+2.4 \pm \ 0.5$ & $+23.1 \pm \ 0.3$ & $-46.7 \pm \ 0.3$ & $-26.6 \pm \ 0.3$ & $+3.0 \pm \ 0.3$ & ($j$) \\
o(sk)a & $+11.1 \pm \ 0.3$ & $+3.0 \pm \ 0.3$ & $+3.4 \pm \ 0.2$ & $-17.0 \pm \ 0.2$ & $-8.9 \pm \ 0.2$ &$ -4.7 \pm \ 0.2$ & ($g$) \\
a(sk)w & $+21.8 \pm \ 0.9$ & $+2.4 \pm \ 0.5$ & $-8.9 \pm \ 0.2$ & $-4.7 \pm \ 0.2$ & $+14.7 \pm \ 0.8$ & $-28.3 \pm \ 0.8$ & ($h$) \\
w(sk)a & $+21.8 \pm \ 0.9$ & $+2.4 \pm \ 0.5$ & $+14.7 \pm \ 0.8$ & $-28.3 \pm \ 0.8$ & $-8.9 \pm \ 0.2$ & $-4.7 \pm \ 0.2$ & ($g$) \\
s(ko)a & $ -4.7 \pm \ 0.2 $& $-14.3 \pm \ 0.3$ & $-17.0 \pm \ 0.2$ & $ -6.6 \pm \ 0.2$ & $+3.0 \pm \ 0.3$ & $-26.6  \pm \ 0.3 $& ($c$)\\
a(ko)s & $ -4.7 \pm \ 0.2$ & $-14.3 \pm \ 0.3 $&$ +3.0 \pm \ 0.3$ & $-26.6 \pm \ 0.3$ & $-17.0 \pm \ 0.2$ &$ -6.6 \pm \ 0.2 $& ($b$)\\
\end{tabular}
\end{ruledtabular}
\captionsetup{justification=Justified,width=\textwidth} 
\caption{\label{tab:spreading parameter} {\bf Fluid combinations, spreading parameters and wetting states.} 
The key d(12)o indicates the combination of droplet/bubble fluid (d) the two lubricants (12) and the outer fluid (o).
The working fluids used are water (w), Krytox (k), olive oil (o) silicone oil (s) and air (a).
Spreading parameters are reported in~mN\,m$^{-1}$.}
\end{table*}

To illustrate this idea, let us focus on state ($j$), where the droplet has double cloaking and the lubricants switch cloaking roles underneath and outside of the droplet, i.e., $\stj$. 
The droplet's surface comprises three interfaces: the interface between the droplet and lubricant~1~(d1), that  between lubricants~1 and~2~(12), and that between the outer~fluid and lubricant~2~(o2). 
The effective interfacial tension is thus $\gamma_{\rm do}^{\mathrm{eff}}=\gamma_{\rm d1}+\gamma_{\rm 12}+\gamma_{\rm o2}$. 
Underneath the droplet lubricant 1 cloaks lubricant 2. 
Hence, one has a continuous interface between the droplet and lubricant 1, and a discontinuous interface between lubricants 1 and 2 (because of the pattern). 
If $f$ is the fraction of lubricant 2, then the effective interfacial tension between the droplet and the PaLS is $\gamma_{\rm d,PaLS}^{\mathrm{eff}}=\gamma_{\rm d1} + f\gamma_{12}$. 
Following a similar argument, the effective interfacial tension between the outer fluid and the PaLS is  $\gamma_{\rm o,PaLS}^{\mathrm{eff}}=\gamma_{\rm o2} + (1-f) \gamma_{\rm 12}$. 
Imposing balance between the 3 interfacial tensions on the plane of the PaLS we arrive at a new law for the apparent contact angle 
\begin{eqnarray}
\label{eq:CassieLiquid}
\cos \theta_{\rm app} & = & \frac{\gamma_{\rm o,PaLS}^{\mathrm{eff}}-\gamma_{\rm d,PaLS}^{\mathrm{eff}}}{\gamma_{\rm do}^{\mathrm{eff}}}\\
& = & \frac{\gamma_{\rm o2}-\gamma_{\rm d1}+(1-2f)\gamma_{12}}{{\gamma_{\rm d1}+\gamma_{12}+\gamma_{\rm o2}}}.\nonumber
\end{eqnarray}

The same approach can be used to derive expressions for the apparent contact angle in each wetting state, which we report in Table~\ref{tab:table equation}. 
%
These expressions demonstrate the complexity and richness of the wettability of the PaLS.
For 8 of the wetting states, the apparent contact angle depends on the surface pattern fraction, $f$.  
Notably, for states ($d$) and ($i$) the apparent contact angle is insensitive to $f$. 
This occurs when the same lubricant cloaks underneath and outside of the droplet, and  
hence the droplet does not interact with the liquid pattern. 
As a consequence, the apparent contact angle matches that observed for droplets resting on the corresponding single-lubricant surface, making these configurations degenerate in $\theta_{\rm app}$. 
One may be tempted to argue that the same should occur for states ($e$) and ($j$) as the droplet only interacts with lubricant 1 underneath and the outer phase only interacts with with lubricant 2 outside. 
However, this disregards the contribution of the interfacial tensions between lubricants 1 and 2, which are weighted differently by the surface pattern fraction in each region.

Eq.~\ref{eq:CassieLiquid} can also be used to understand the failure of the classical Cassie-Baxter law, Eq.~(\ref{eq:CassieBaxter}), in describing the wettability of PaLS. 
For a given wetting state, Eq.~(\ref{eq:CassieLiquid}) can be recast as a surface averaged law,
$\cos \theta_{\rm app} = \cos \theta_{\rm app}|_{f=0} + f\left(\cos \theta_{\rm app}|_{f=1}-\cos\theta_{\rm app}|_{f=0}\right).$     
Using state ($j$) as an example yields $\cos \theta_{\rm app}|_{f=0}=(\gamma_{\rm o2}-\gamma_{\rm d1}+\gamma_{12})/({\gamma_{\rm d1}+\gamma_{12}+\gamma_{\rm o2}})$ and $\cos \theta_{\rm app}|_{f=1}=(\gamma_{\rm o2}-\gamma_{\rm d1}-\gamma_{12})/({\gamma_{\rm d1}+\gamma_{12}+\gamma_{\rm o2}})$, showing that the limiting apparent contact angles 
are affected by both lubricants, and not just by one.   
 
\subsection{Comparison Between Theory, Simulations and Experiments}



To test the prediction of Eq.~(\ref{eq:CassieLiquid}), we carried out further experiments of liquid droplets and bubbles on patterned liquid surfaces using combinations of water, silicone oil, olive oil, Krytox and air as working fluids. 
The choice of these fluids allows us to access five of the ten possible wetting states, capturing different cloaking configurations. 
The corresponding spreading coefficients and wetting states are summarized in Table~\ref{tab:spreading parameter}. 
For example, a droplet of olive oil on a patterned liquid surface formed by stripes of silicone oil and Krytox surrounded by air (marked `o(sk)a' in the table) has spreading coefficients that map to state ($g$). 
For each combination, we varied the surface fraction and measured the corresponding apparent contact angle. 
We also carried out numerical simulations for each wetting state, focusing on controlling the apparent contact angle by either varying the surface fraction or the interfacial tensions.

\begin{figure}
    \centering
    \includegraphics[width=0.45 \textwidth]{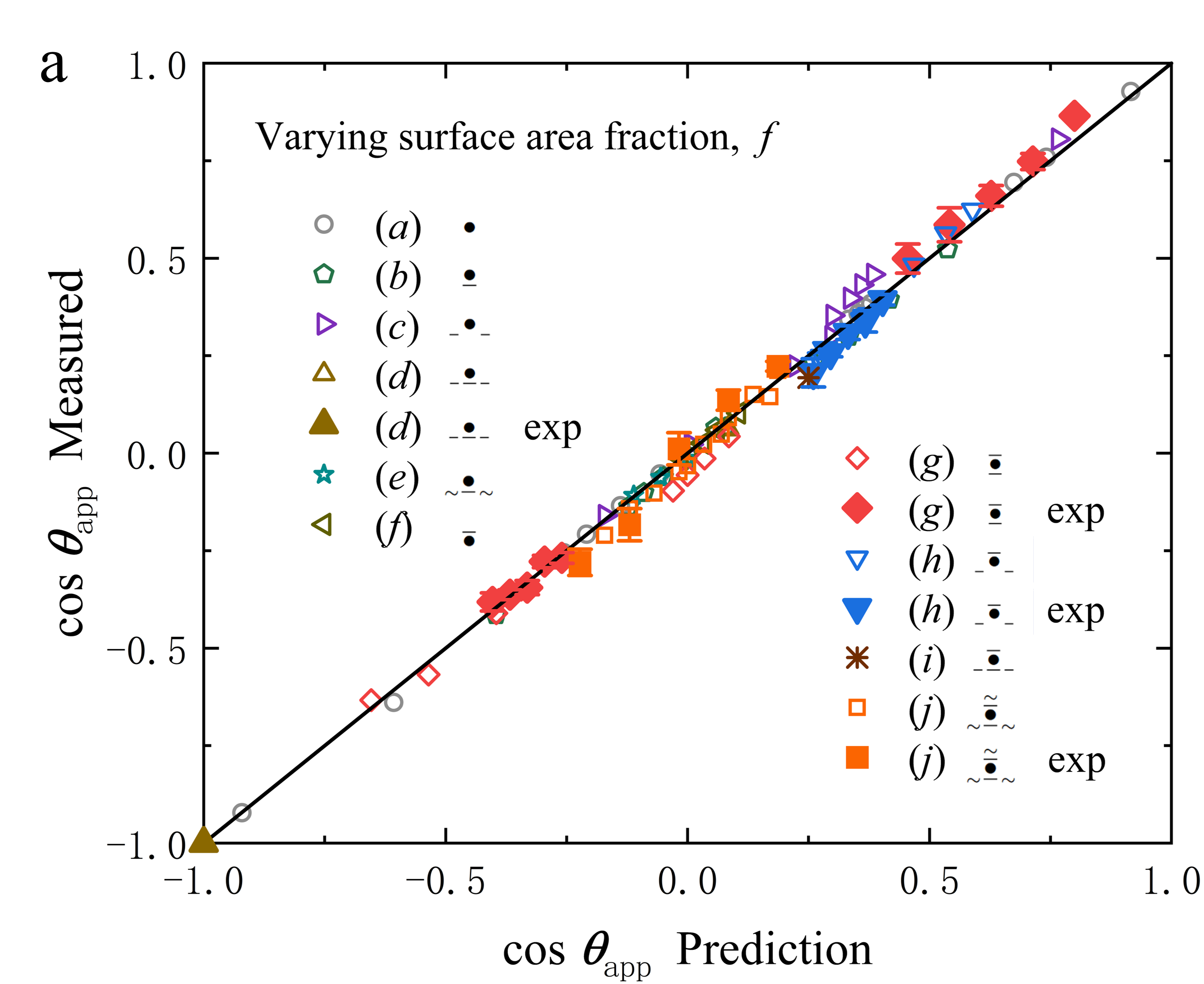}
    \includegraphics[width=0.45 \textwidth]{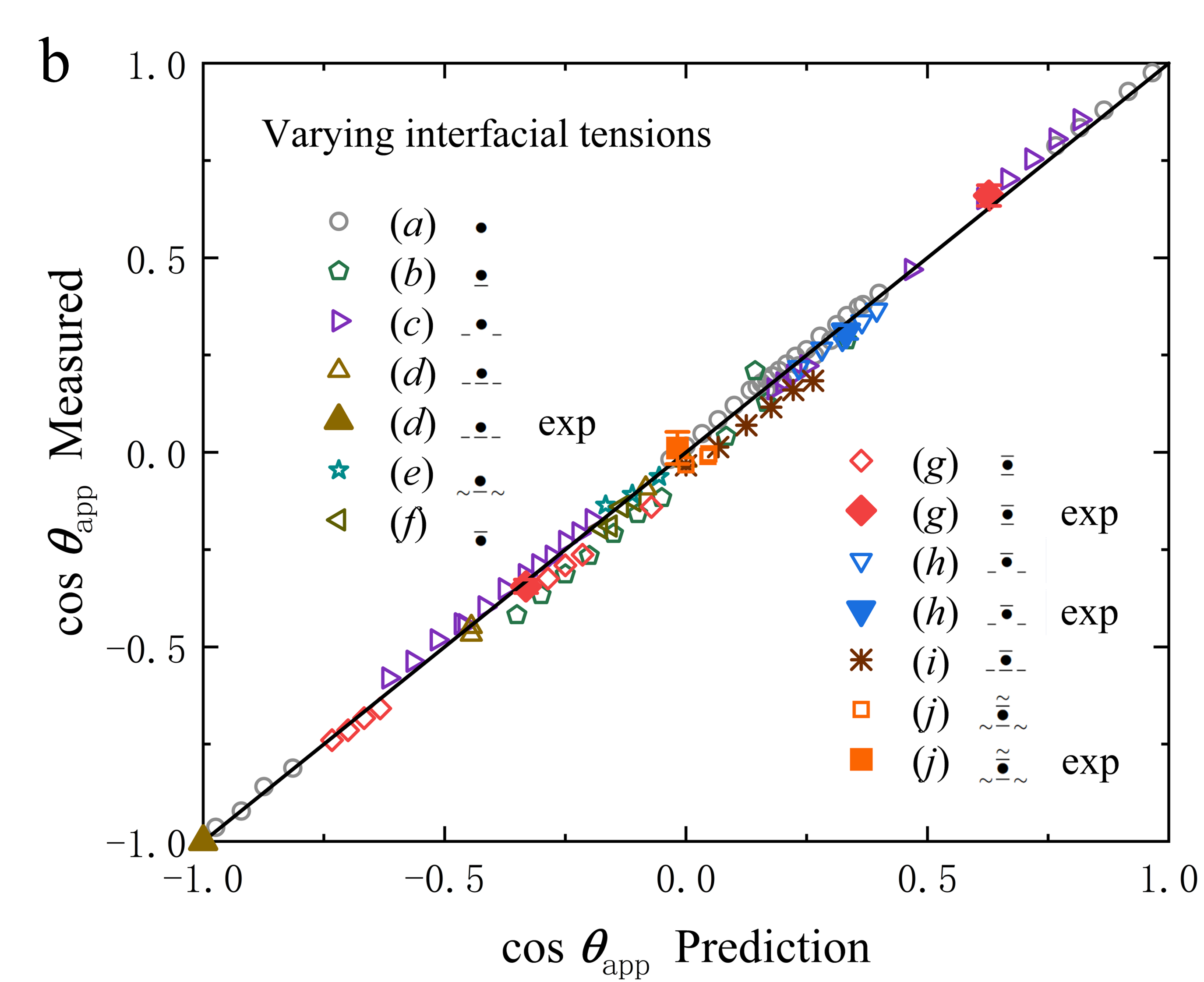}
    \captionsetup{justification=Justified,width=\columnwidth}
    \caption{{\bf Measured vs predicted apparent contact angles on PaLS.} {\bf a} Data from experiments and simulations for different fluid combinations and varying surface pattern fraction. {\bf b} Data from experiments and simulations at fixed surface fraction $f=0.5$ and varying interfacial tensions. States and configurations are indicated in the legend.   
    }
    \label{fig:Fig4} 
\end{figure}

The results are shown in Fig.~\ref{fig:Fig4}.
The panels show measured versus predicted values of the apparent contact angle for droplets placed on surfaces with different surface pattern fraction, $f$ (Fig.~\ref{fig:Fig4}a), or with different interfacial tensions for a fixed surface fraction $f=0.5$ (Fig.~\ref{fig:Fig4}b). 
The predicted apparent angles are obtained by first determining the spreading coefficients based on the interfacial tensions (which are either measured in the experiments or provided as input parameters in the numerical simulations).
This allows us to identify the corresponding wetting state and to select the appropriate expression for $\theta_{\rm app}$ from Table~\ref{tab:spreading parameter}.
In total, we tested {\color{black} 190 droplet configurations} finding an excellent agreement with the theoretical prediction.
Small disparities can be attributed mainly to the fact that the lubricant wetting ridge is small but still finite in size. 
The wetting ridge can exert forces on the interface of droplets, making the interface curvature adjust in order to satisfy the balance of Laplace pressure across the interfaces of the ridges and droplets, and thus affecting  the apparent contact angle~\cite{Semprebon2017SoftMatter}. We discuss the effect of the wetting ridge size on the apparent contact angle in the Supplementary Materials. 
Importantly, the results in Fig.~\ref{fig:Fig4} underscore the ability to tailor droplet contact angles on PaLS over the full range of wettability ($0-180^\circ$) by varying the lubricant fraction and selecting suitable liquids. 

\begin{figure*}
    \centering
    \includegraphics[width=0.9\textwidth]{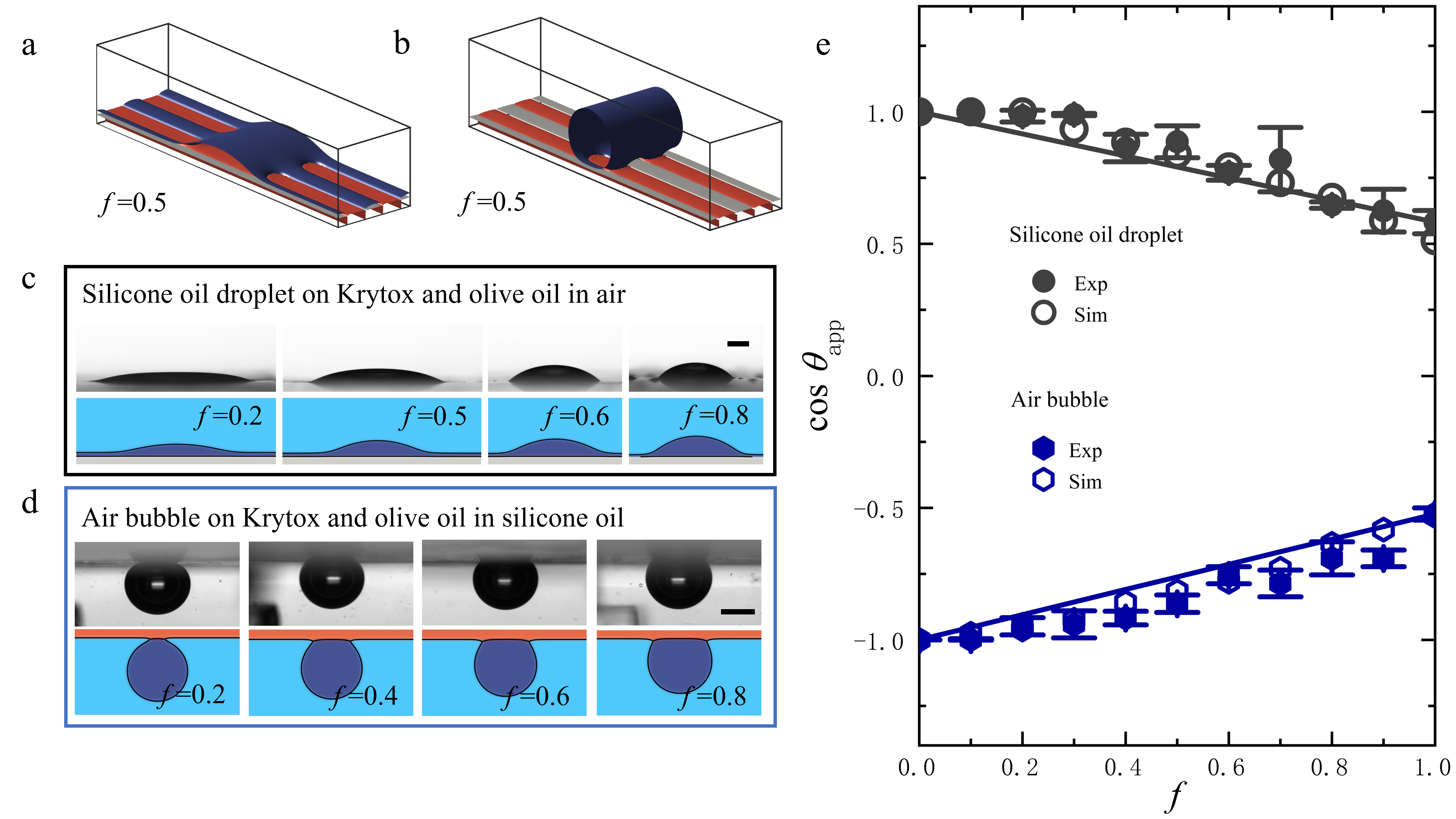}
    \captionsetup{justification=Justified,width=\textwidth} 
    \caption{{\label{fig:Fig5} \bf Controlling spreading and dewetting using PaLS.} {\bf a} A fluid that would otherwise spread into a thin film on a lubricant layer is stablised into a droplet shape by the second lubricant. {\bf b} A fluid that would otherwise detach from a lubricant layer is forced to attach by the second lubricant. {\bf c} and {\bf d}, Experimental observation of the configurations of {\bf a} and {\bf b}. The bottom rows show cross-sections of the droplet obtained from numerical simulations. {\bf e} Cosine of the apparent contact angle vs surface pattern fraction for the fluid combinations of {\bf c} and {\bf d}. The solid line corresponds to the theoretical prediction (see text). The scale bars in {\bf c} and {\bf d} are 1 mm.}
\end{figure*}

\section{Conclusions}
Here we have developed a new method to create patterned liquid surfaces with distinct micron-sized lubricant domains.  
PaLS combine the ultrasmoothness of the liquid lubricants with the structure provided by the surface pattern, thus unlocking a new mechanism to adjust the wettability of a lubricant-infused surface over the full range of apparent contact angles.
We have shown that the lubricant pattern introduces anisotropy in the surface wettability, including capillary adhesion forces that can be used to keep a droplet in place. 
%
These forces arise purely from the lubricant domains and can be adjusted by a rational design of the pattern geometry.
The richness of multiphase equilibria on PaLS is illustrated by the variety of equilibrium configurations of droplets on these surfaces.  
Moreover, we have provided a systematic approach to identify and study the equilibrium states based on the spreading coefficients of the specific multiphase fluid system and have developed a rational approach to predict the wettability of the surface via the apparent contact angle.  

So far, we have explored different possible lubricant cloaking combinations while limiting our study to situations where the droplet shows partial wetting separately on both lubricants.  
Further equilibrium states are possible if we relax this constraint. For instance, one may consider situations where the droplet either completely spreads on, or completely detaches from, one of the two lubricants.
A fluid that completely spreads on lubricant 2 ($S_{\rm d2o}>0$) but not on lubricant 1 ($S_{\rm d1o}<0$) can be stabilized into a droplet shape on a patterned liquid surface as shown in Fig.~\ref{fig:Fig5}a thus preventing complete spreading.  
Likewise, a fluid that completely dewets from lubricant 2 ($S_{\rm o2d}>0$) but not from lubricant 1 ($S_{\rm o1d}<0$) can be forced to attach to the surface using the same approach (Fig.~\ref{fig:Fig5}b).
These situations can be realized experimentally, as shown in Figs.~\ref{fig:Fig5}c and~\ref{fig:Fig5}d.
Furthermore, because the apparent contact angle is controlled by the wettability of both lubricants, it is possible to vary the apparent contact angle over a broad range by changing the surface pattern fraction. 
Following the same approach as before, we find 
%
\begin{eqnarray}
\label{eq:CassieLiquid}
\cos \theta_{\rm app}& = & \frac{f (\gamma_{\rm 2o}-\gamma_{\rm 2d})+(1-f)\gamma_{\rm do}}{{\gamma_{\rm do}}},\nonumber
\end{eqnarray}
which is in excellent agreement with the experimental and simulation data (see Supplementary Materials for values of the surface tensions). 
These results further illustrate how PaLS can be used to control the configuration of droplets and bubbles without the need of direct contact with a solid surface.

The ability to control the wettability of PaLS down to the scale of micron-sized domains offers advantages in practical applications. 
For example, in MEMS and droplet microfluidics technologies, interactions between droplets and solids are often addressed using chemical or topographical patterns, both of which are limited by contact-line pinning~\cite{malinowski2020advances,zeng2024MaterialsTodayPhysics}. 
On PaLS, pinning on the solid is eliminated while the structure of the lubricant domains introduces capillary interactions which can be engineered control the position of a target droplet by creating anchoring points or to induce self-propulsion via a gradient in the pattern geometry. 
In inkjet printing applications~\cite{lohse2022fundamental}, droplet-droplet interactions mediated by the surface are crucial, but are often limited by the quality of the underlying solid substrate. 
Here, PaLS could be used as a primer layer~\cite{wijshoff2018drop} and exploited to to increase the resolution quality of current wet-on-wet printing techniques~\cite{hack2018printing}.

\section{Methods}
\subsection{Surface Preparation}
Pre-cleaned glass slides (CLS294775X25-72EA, Corning) were dip-coated in a silica-based nano-particle suspension (Glaco Mirror Coat Zero, SOFT 99 Corp) at a speed of 1~{mm~s$^{-1}$} using a robot (Ossila LTD). 
A residence time of 30 seconds was allowed before the substrates were withdrawn from the solution at a speed of {0.5~mm~s$^{-1}$}. 
The substrates were left to dry for 3 minutes. 
The dip-coating process was repeated 5 times to ensure that a uniform layer of nano-particles was formed on the surfaces~\cite{guan2017drop,luo2017slippery}.  

The substrates were primed with hexamethyldisilazane (HMDS) vapor for 10 minutes. 
A sacrificial layer of photoresist (SPR~350, Megaposit) was deposited on the substrates using a spin coater (POLOS200, SPS POLOS) at a speed of 4000~rpm. 
The coated substrates were baked at $90^\circ{\rm C}$ for 90 seconds. 
The photoresist was exposed to UV light at 44~$\rm{mJ~cm^{-2}}$ to create the desired pattern using a microwriter (ML3, Durham Magneto Optics). 
The substrates were rinsed with a developer solution (MF-26A, Kayaku Advanced Materials) to remove the exposed regions of the photoresist.

The substrates were treated with oxygen plasma at 200~W for 2 minutes. 
The surfaces were then fluorosilanized via chemical vapor deposition (CVD) using Trichloro(1H,1H,2H,2H-perfluorooctyl)silane (No. 448931, Sigma-Aldrich) before the sacrificial photoresist was removed by rinsing with isopropanol (Propan-2-ol, $\geq 99.8\%$, Sigma-Aldrich) for 30 seconds. 
The substrates were coated with the target lubricants using sequential dip-coating at a withdrawal speed of 0.01~mm~s$^{-1}$.
%
%

%

\subsection{Simulation Model}

To determine the stable wetting states of droplets on PaLS, numerical simulations of a phase-field model were conducted by minimizing the total free-energy of the four fluid phases, as given in Eq.~\ref{total free energy}. Here we describe the different contributions to the free-energy. The simulation setup is discussed in the Supplementary Material. 

The bulk free-energy $E_{\text{bulk}}$ is defined as:
\begin{equation}
    \begin{split}
    E_{\text{bulk}} = &\int\left(\sum_{i=1}^{N-1}\sum_{j=i+1}^{N}k_{ij}[f(c_i)+f(c_j)+f(c_i+c_j)]\right. \\
    &\left.+ \lambda\sum_{i=1}^{N-2}\sum_{j=i+1}^{N-1}\sum_{k=j+1}^{N}c_i^2c_j^2c_k^2\right) \, dV,
    \end{split}
    \label{bulk free energy}
\end{equation}
in which $N=4$ is the total number of the fluid phases and $V$ is the total volume occupied by them. 
The first term allows for the coexistence of the 4 fluid phases. 
Here, the local concentration of the $i$-th fluid phase, $c_i$, can be regarded as an order parameter which varies from 0 to 1. 
The function $f(c_i)=c_i^2(c_i-1)^2$ is a double-well potential whose minima, located at $c_i=0$ and $c_i=1$, characterize the pure phases of the mixture. $k_{ij}$ are numerical prefactors that are related to the interfacial tensions, see Eq. \eqref{eq:ST}. 
Following Boyer and Lapuerta \cite{boyer_study_2006}, the second term in Eq.~(\ref{bulk free energy}) is introduced to avoid the instability of the phase field when one phase spreads over another. 
In this term, $\lambda$ is a positive stabilization parameter, which we set to $\lambda=50$ in lattice unit (l.u.).

The interfacial free energy $E_{\text{surf}}$ is defined by:
\begin{equation}
    \begin{split}
    E_{\text{surf}} = &\int \left(\sum_{i=1}^{N-1} \frac{k_{ii}^{'}}{2} \lvert \nabla{c_i} \rvert^2+ \right.
    \\
    & \left. \sum_{i=1}^{N-2}\sum_{j=i+1}^{N-1}\frac{k_{ij}^{'}}{2}\lvert \nabla{(c_i+c_j)} \rvert^2 \right) \, dV.
    \end{split}
\end{equation}
In equilibrium, it is possible to derive expressions for the interfacial tensions between the different phases, $\gamma_{ij}$, from $E_{\rm bulk}$ and $E_{\rm surf}$. They are related to the $k$-prefactors by 
\begin{equation}
\label{eq:ST}
\gamma_{ij}=\frac{\alpha}{3}\sum_{m=1}^{N-1}\sum_{n=m+1}^{N}w_{mn}k_{mn},
\end{equation}
where the weights $w_{mn}$ are defined by
\begin{equation}
    w_{mn} = \begin{cases}
               1 & \text{if } m=i \text{ or } j \text{ for any }n, \\
               1 & \text{if } n=i \text{ or } j \text{ for any }m,\\
               0 & \text{if otherwise},  \\
           \end{cases}
\end{equation}
and $\alpha$ is a measure of the width of diffuse interface, set to be 1.0 l.u.. 
%
The parameters \( k'_{ij} \) are associated with the spreading parameters \( S_{Nij} \)  via:
\begin{equation}
   k'_{ij}= 
\begin{cases}
    -3\alpha S_{Nij} & \text{if } i \neq j\\
    3\alpha \left( 2\gamma_{iN} + \sum_{m=1, m\neq i}^{N-1} S_{Nim} \right ),              & \text{if } i = j.
\end{cases}
\end{equation}
In our simulations, the surface tension parameters are chosen to mimic the liquids used in the experiments (see Supplementary Materials), or they are systematically varied to access different wetting states. 



In the experiments, the micro surface patterns make one set of regions of the solid substrate energetically attractive to the first lubricant liquid, and the other regions attractive to the second lubricant liquid. To model this spatial confinement of each lubricant liquid phase in the simulations, we employ the confining free energy, 
\begin{equation}
    E_{\text{conf}}=\int_{\Omega_{\text{lub}}}  \Psi_{\text{conf}} \, dV,
    \label{total conf free energy}
\end{equation}
where, $\Omega_{\text{lub}}$ represents the lubricant region with a thickness of 10 l.u. from the bottom of the simulation domain.
Without loss of generality, we now index the two lubricants as phases 1 and 2, the droplet as phase d, and the surrounding fluid as phase o. In regions of the space where lubricant 1 is confined, $\Psi_{\rm{conf}}$ takes the form:
\begin{equation}
   \Psi_{\rm{conf}}=C_{\rm{conf}} \left[ ( c_d-1)^3(3c_d+1) + c_1^3(3c_1-4) \right], 
   \label{eqn:psi_conf_2}
\end{equation}
whereas in regions of space where lubricant 2 is confined,
\begin{equation}
   \Psi_{\rm{conf}}=C_{\rm{conf}} \left[ ( c_d-1)^3(3c_d+1) + c_2^3(3c_2-4) \right],
   \label{eqn:psi_conf_3}
\end{equation}
where $C_{\rm{conf}}$ is the confining potential strength, set at $C_{\rm{conf}}=1$ in l.u.. 
The first terms in both Eqs. \eqref{eqn:psi_conf_2} and \eqref{eqn:psi_conf_3} make the regions of space where the lubricant is confined repulsive to the droplet phase. The second terms in each equation make the region attractive to lubricants 1 and 2, respectively.

The ensemble free energy, $E_{\text{ens}} = E_{\rm drop} + E_{\rm lub}$, accounts for the constant-volume ensemble for the droplet and constant-pressure ensemble for the lubricants. 
The constant volume ensemble for the droplet is enforced by the term
\begin{equation}
    E_{\text{drop}}=k_{\rm vol}(V_{\rm drop}-V_{0})^2.
    \label{volume constrain}
\end{equation}
Here, $k_{\text{vol}}=0.001$ (in l.u.) is a constant prefactor controlling the strength of the constraint, $V_{\rm drop}$ is the measured volume of droplet phase in the domain, and $V_{0}$ is the target volume of the droplet. 
%
%
For the lubricant liquids, we apply the pressure constraint (instead of imposing a constraint on their volume) to ensure that a sufficient amount of lubricants is available to fully spread on the droplet and/or the other interfaces. 
The pressure constraint is expressed as follows:
\begin{equation}
    E_{\text{lub}}=-p_0V_{\text{lub}},
    \label{pressure constrain}
\end{equation}
where $p_0<0$ is the pressure of the lubricant liquids. We use a range of $-0.05<p_0<-0.01$ in this work.
A small absolute value of $p_0$ indicates less work to increase the volume of the lubricant layer at the expense of additional deformation of its interface. This leads to larger ridge sizes, especially in the cases where one lubricant fully wets the surface of the lubricant layer outside of the droplet and also cloaks the droplet interface. 

\begin{acknowledgments}
We thank Michele Pelizzari for useful discusions on the surface preparation. We acknowledge funding from the Leverhulme Trust (Research Project Grant RPG-2022-140) and UKRI Engineering and Physical Sciences Research Council (EP/V034154/2).
\end{acknowledgments}

\section*{Author Contributions}
R.L.-A., H.K., G.M. and G.G.W conceptualized the research. 
X.Z., J.P. and H.K. developed the simulation method. 
X.Z. performed the simulations and analyzed the simulation data supported by J.P., R.L.-A. and H.K. 
H.Z., G.M. and G.G.W. conceptualized and developed the surface preparation method.
H.Z., G.G.W. and R.L.-A. conceptualized the experimental setup. 
H.Z. performed the experiments and analyzed the
experimental data with support from G.G.W, R.L.-A. and G.M. 
H.Z and X.Z. produced the figures. 
All authors contributed to interpreting the results.
R.L.-A. and H.K. wrote the paper with contributions from all authors. 

\section*{Data availability}
The datasets generated during and/or analyzed during the current study are available from the corresponding authors upon reasonable request.

\section*{Code availability}
The phase field energy minimisation code used in the current study is available from the corresponding authors on reasonable request.



\end{document}



\title{Multiple Equilibria Enables Tunable Wetting of Droplets on Patterned Liquid Surfaces\\
Supplementary Materials}

\author{Xitong Zhang}
\thanks{These authors contribute equally.}
\affiliation{Institute for Multiscale Thermofluids, School of Engineering, The University of Edinburgh, King's Buildings, Mayfield Road, Edinburgh EH9 3FD, U.K.}
\affiliation{Department of Physics, Durham University, South Road, Durham DH1 3LE, U.K.}
\author{Hongyu Zhao}
\thanks{These authors contribute equally.}
\affiliation{Institute for Multiscale Thermofluids, School of Engineering, The University of Edinburgh, King's Buildings, Mayfield Road, Edinburgh EH9 3FD, U.K.}
\author{Jack R. Panter}
\affiliation{School of Engineering, University of East Anglia, Norwich, Norfolk NR4 7TJ, UK.}
\author{Glen McHale}
\email{glen.mchale@ed.ac.uk}
\affiliation{Institute for Multiscale Thermofluids, School of Engineering, The University of Edinburgh, King's Buildings, Mayfield Road, Edinburgh EH9 3FD, U.K.}
\author{Gary G. Wells}
\email{gary.wells@ed.ac.uk}
\affiliation{Institute for Multiscale Thermofluids, School of Engineering, The University of Edinburgh, King's Buildings, Mayfield Road, Edinburgh EH9 3FD, U.K.}
\author{Rodrigo Ledesma-Aguilar}
\email{rodrigo.ledesma@ed.ac.uk}
\affiliation{Institute for Multiscale Thermofluids, School of Engineering, The University of Edinburgh, King's Buildings, Mayfield Road, Edinburgh EH9 3FD, U.K.}
\author{Halim Kusumaatmaja}
\email{halim.kusumaatmaja@ed.ac.uk}
\affiliation{Institute for Multiscale Thermofluids, School of Engineering, The University of Edinburgh, King's Buildings, Mayfield Road, Edinburgh EH9 3FD, U.K.}
\affiliation{Department of Physics, Durham University, South Road, Durham DH1 3LE, U.K.}

\date{\today}

\begin{abstract}

\end{abstract}

\maketitle

\renewcommand{\thetable}{A\arabic{table}} 
\renewcommand{\thefigure}{A\arabic{figure}} 
\renewcommand{\theequation}{A\arabic{equation}} 
\renewcommand{\thefootnote}{\Alph{footnote}} 

\section{Supplementary Experimental Methods}

\subsection{Interfacial Tension Measurements}


Interfacial tensions were measured using the pendant droplet method.
%
A droplet (fluid $i$) was dosed using a syringe pump into the surrounding phase (fluid $j$) in a transparent cuvette container. 
%
%
The droplet was recorded by a CCD camera and its shape was processed by ImageJ using a pendant droplet analysis algorithm~\cite{daerr2016pendent_drop}.
The interfacial tension, $\gamma_{ik}$, is obtained from the balance between the capillary and gravitational forces,
\begin{equation}
    \gamma_{ij}=\frac{\Delta \rho_{ij} g R}{\beta},
    \label{surface tension}
\end{equation}
where $\Delta \rho_{ij}$ is the density difference between the fluids, $g$ is the acceleration due to gravity, $R$ is the radius of curvature at the tip of the droplet and $\beta$ is a factor derived from the shape of the droplet.
For each fluid-fluid pair, the interfacial tension was determined by averaging five independent measurements. The standard error was obtained by the error propagation based on Eq. \ref{surface tension}. 
The measured values for each fluid-fluid pair are reported in Table \ref{tab:interfacial tension}. 

\begin{table*}[t!]
\begin{ruledtabular}
\begin{tabular}{cccccccccc}
$\gamma_{\rm wk}$ & $\gamma_{\rm wo}$ & $\gamma_{\rm ws}$ & $\gamma_{\rm wa}$ & $\gamma_{\rm ko}$ & $\gamma_{\rm ks}$ & $\gamma_{\rm ka}$ & $\gamma_{\rm os}$ & $\gamma_{\rm oa}$ & $\gamma_{\rm sa}$ \\
\hline
$52.5 \pm 0.2$ & $17.6 \pm 0.2$ & $30.0  \pm 0.5$ & $71.7 \pm 0.4$ & $11.8 \pm 0.1$ & $6.8 \pm 0.1$ & $16.8 \pm 0.1$ & $1.6 \pm 0.1$ & $31.6 \pm 0.3$ & $18.9 \pm 0.1$ \\
\end{tabular}
\end{ruledtabular}
\captionsetup{justification=Justified,width=\textwidth} 
\caption{\label{tab:interfacial tension} {\bf  Measured interfacial tensions between fluids used in experiments (mN\,m$^{\boldsymbol{-}\boldsymbol{1}}$).
} 
The subscripts stand for water (w), Krytox (k), olive oil (o) silicone oil (s) and air (a).
}
\end{table*}

\subsection{Apparent Contact Angle Measurements}
The apparent contact angle of a droplet in contact with a lubricant-infused surface is affected by the presence of the wetting ridge formed by the lubricant and the droplet \cite{guan_evaporation_2015, Semprebon2017SoftMatter}.
We consider two natural geometric choices to define the apparent contact angle: the first is determined by the slope of the droplet–gas interface at the intersection between the wetting ridge and the droplet, denoted $\theta_{\text{app}}^*$; the second is determined by the slope of the virtual droplet–gas interface extrapolated to the position of the surface of the lubricant layer, denoted $\theta_{\text{app}}$.
When there is little excess lubricant and the wetting ridge is small, as is the case in the experiments, we expect that there is no significant difference between $\theta_{\text{app}}^*$ and $\theta_{\text{app}}$. 
Therefore, we measure the apparent contact angle by fitting the interface profile of the droplet to a third-degree polynomial and extrapolating its slope to the baseline of the surface. 
%
Measurements were repeated five times and averaged. 
The error is reported as the standard deviation of the sample. 

\subsection{Sliding Angle Measurements}
%
A sample PaLS made of Krytox and olive oil stripes was placed on a platform connected to a rotating stage (Stepper Motor Rotation Mount, Thorlabs). The initial orientation of the stage was parallel to the horizontal. 
A 10-$\rm{\mu L}$ droplet was gently placed on the PaLS and allowed to equilibrate.
The surface was then tilted at a speed of 0.02$\rm{^\circ~s^{-1}}$. 
The tilting angle at which the contact line of the droplet started to move was recorded as the sliding angle.
Due to the anisotropy of the stripe pattern, the experiment was carried out by tilting the surface along the parrallel and the perpendicular directions to the stripes separately.
The sliding angle for a droplet on the corresponding single-lubricant infused surface was measured for comparison.
The standard deviation was given from the measurement of four independent experiments.

\section{Supplementary Simulation Methods}

\subsection{Cylindrical Droplet Simulation}
\begin{figure}
    \centering
    \includegraphics[width=0.6\columnwidth]{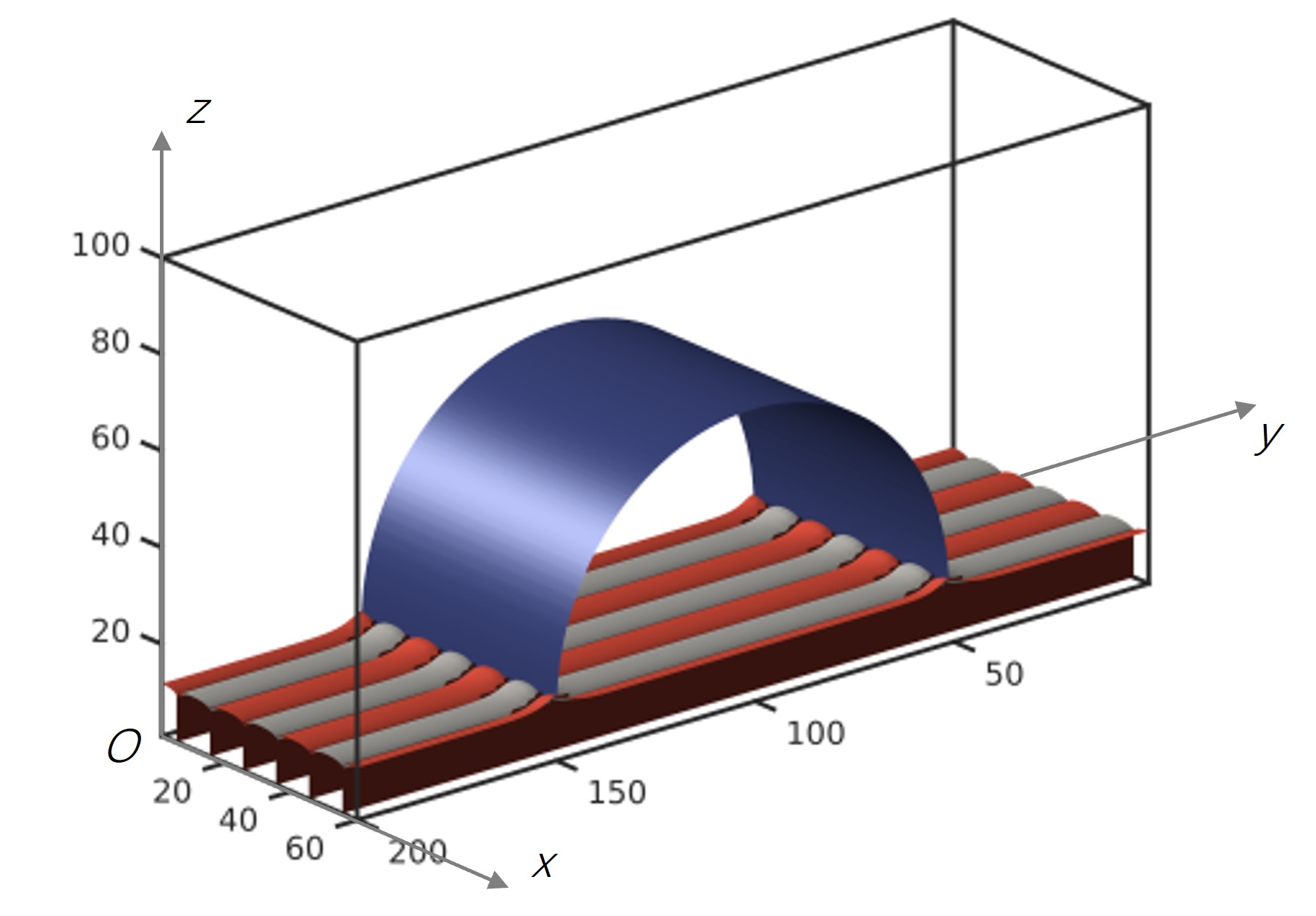}
    \caption{Sketch of computational domain in simulation. For clear observation purposes, only the surface of the droplet is shown, and the interior of the droplet is set to a transparent color.}
    \label{fig:domain}
\end{figure}

In this work we primarily simulate cylindrical droplets on PaLS since we are interested in the apparent contact angle parallel to the stripes. Figure~\ref{fig:domain} shows a schematic of the computational simulation domain. 
The simulation box has a rectangular cross-section of dimensions \( L_y = 200 \) and \( L_z = 100 \), both in lattice units (l.u.). 
The width of the domain, \( L_x \), is fixed depending on the width of the lubricant stripes. 
A minimum of three periods of the lubricant pattern are resolved in the simulations, with a minimum stripe width of 10 l.u.  
The top and bottom boundaries of the domain are modeled as solid walls, while mirror boundary conditions are applied for the remaining boundaries. 
The lubricant layer is initially subjected to a pressure constraint and undergoes a free energy minimization process. 
Following this step, the droplet is initialized as a half-cylinder with a constant volume constraint. We typically set the droplet volume to be at least 12\% of the volume of the domain.
%
Subsequently, we minimise the free energy of the entire system. 

\begin{table*} [t!]
\begin{ruledtabular}
\begin{tabular}{crrrrrrc}
$\rm d(12)o$ & $\gamma_{\rm d1}  $& $\gamma_{\rm d2}$ & $\gamma_{\rm do}$ & $\gamma_{\rm 12}$  & $\gamma_{\rm 1o}$ & $\gamma_{\rm 2o}$ & Wetting state\\
\hline
w(ok)a & $1.780$ & $0.596$ & $2.432$ & $0.400$ & $0.568$ & $1.072$ & ($j$) \\
o(sk)a & $0.200$ & $1.400$ & $2.900$ & $0.800$ & $2.350$ &$2.050$ & ($g$) \\
a(sk)w & $0.641$ & $0.569$ & $2.432$ & $0.231$ & $1.050$ & $1.780$ & ($h$) \\
w(sk)a & $1.050$ & $1.780$ & $2.432$ & $0.231$ & $0.641$ & $0.569$ & ($g$) \\
s(ko)a & $0.360$ & $0.085$ & $1.000$ & $0.624$ & $0.889$ & $1.600$& ($c$)\\
a(ko)s & $0.889$ & $1.600 $&$1.000$ & $0.624$ & $0.360$ &$ 0.085 $& ($b$)\\
\end{tabular}
\end{ruledtabular}
\captionsetup{justification=Justified,width=\textwidth} 
\caption{\label{tab:sim parameters for exps} {\bf Interfacial tensions used in simulations for the corresponding experimental cases.} 
The interfacial tensions are in lattice units  (l.u.).
The key d(12)o indicates the combination of droplet/bubble fluid (d), the two lubricants (1 and 2), and the outer fluid (o).
The working fluids used are water (w), Krytox (k), olive oil (o) silicone oil (s) and air (a).
}
\end{table*}
Table~\ref{tab:sim parameters for exps} lists the interfacial tensions used in the simulations that are compared with the corresponding experimental cases in Figs. 2, 4, and 5 of the main paper. In the simulations, the results depend on the ratio of the interfacial tensions, rather than their absolute values.

\subsection{Spherical Droplet (Full 3D) Simulation}

\begin{figure}
    \centering
    \includegraphics[width=0.8\columnwidth]{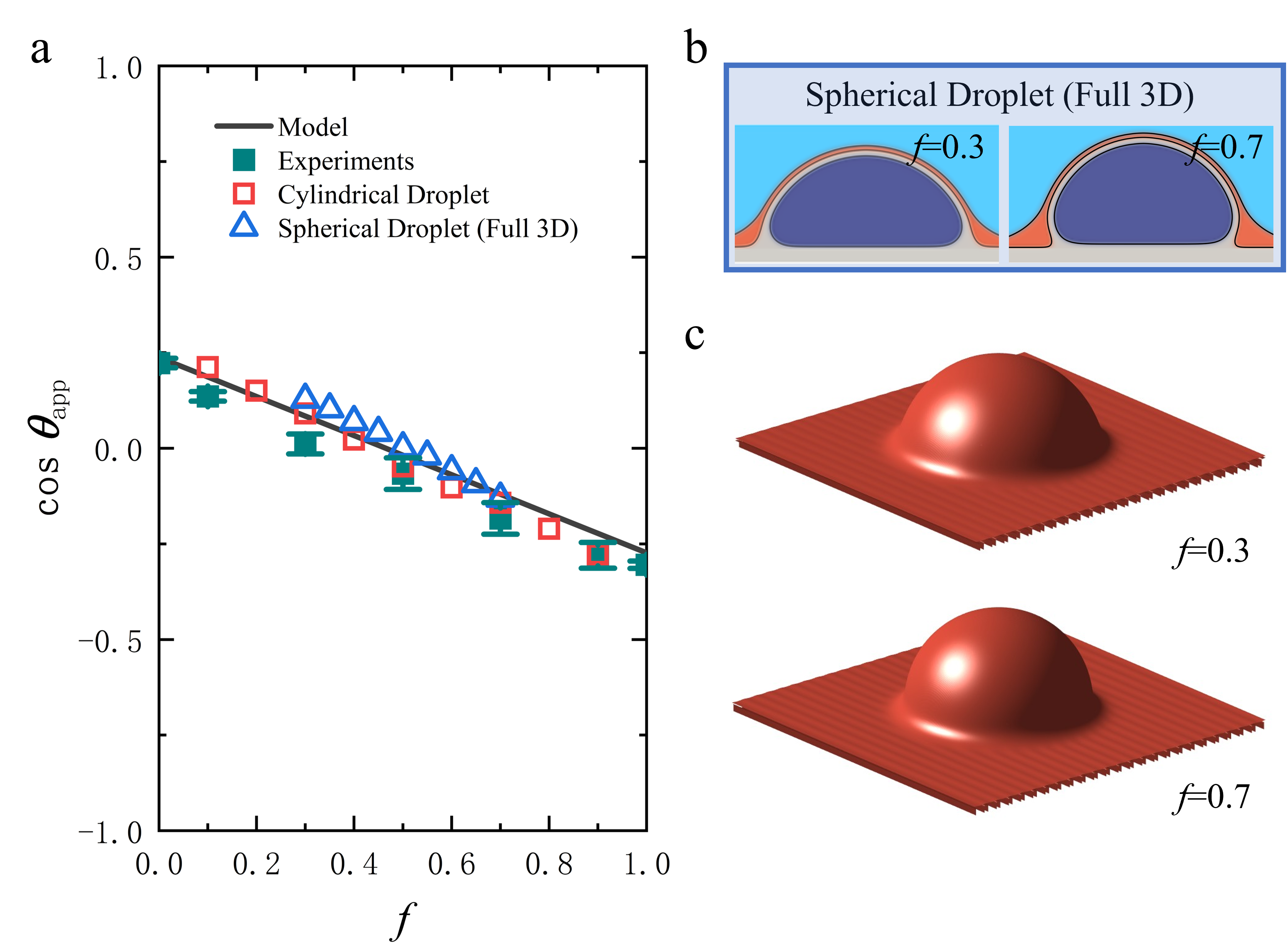}
    \captionsetup{justification=justified, singlelinecheck=false}
    \caption{Comparison of cylindrical and spherical droplet simulations with experimental data for water droplet on olive oil and Krytox PaLS surrounded by air. 
    \textbf{a} Apparent contact angles obtained from both cylindrical and spherical droplet simulations compared to experimental measurements and model prediction, where $f$ represents the Krytox area fraction.
    \textbf{b} Snapshots of the droplet cross-section from full 3D simulations, with Krytox shown in orange and olive oil in gray when $f=0.3$ and $0.7$. 
    \textbf{c} Equilibrium droplet shapes from full 3D simulations for $f=0.3$ and $0.7$.} 
    \label{fig:3D and 2D sim}
\end{figure}

We also conducted a series of full 3D simulations for a water droplet resting on Krytox and olive oil, surrounded by air as the ambient fluid. In this set of simulations, the droplet is initialized as a spherical droplet which then relaxes to its equilibrium configuration. The droplet shape is elongated parallel to the stripes and the apparent contact angle is measured parallel to the stripes. Fig.~\ref{fig:3D and 2D sim}a shows  the apparent contact angles obtained from the cylindrical and spherical droplet simulation setups. Both agree well with those from the experimental measurements and the theory presented in the main paper. This confirms that the cylindrical droplet setup is adequate to capture the apparent contact angle of a droplet on PaLS. Fig.~\ref{fig:3D and 2D sim}b presents snapshots of the central cross-section from the full 3D simulations when the Krytox area fraction is $f = 0.3$ and 0.7. The double-layer cloaking of the droplet corresponds to state ($j$). Fig.~\ref{fig:3D and 2D sim}c shows the 3D droplet morphologies when $f=0.3$ and $0.7$.

\subsection{Effect of Lubricant Ridge Size on the Apparent Contact Angle}

\begin{figure}
    \centering
    \includegraphics[width=\textwidth]{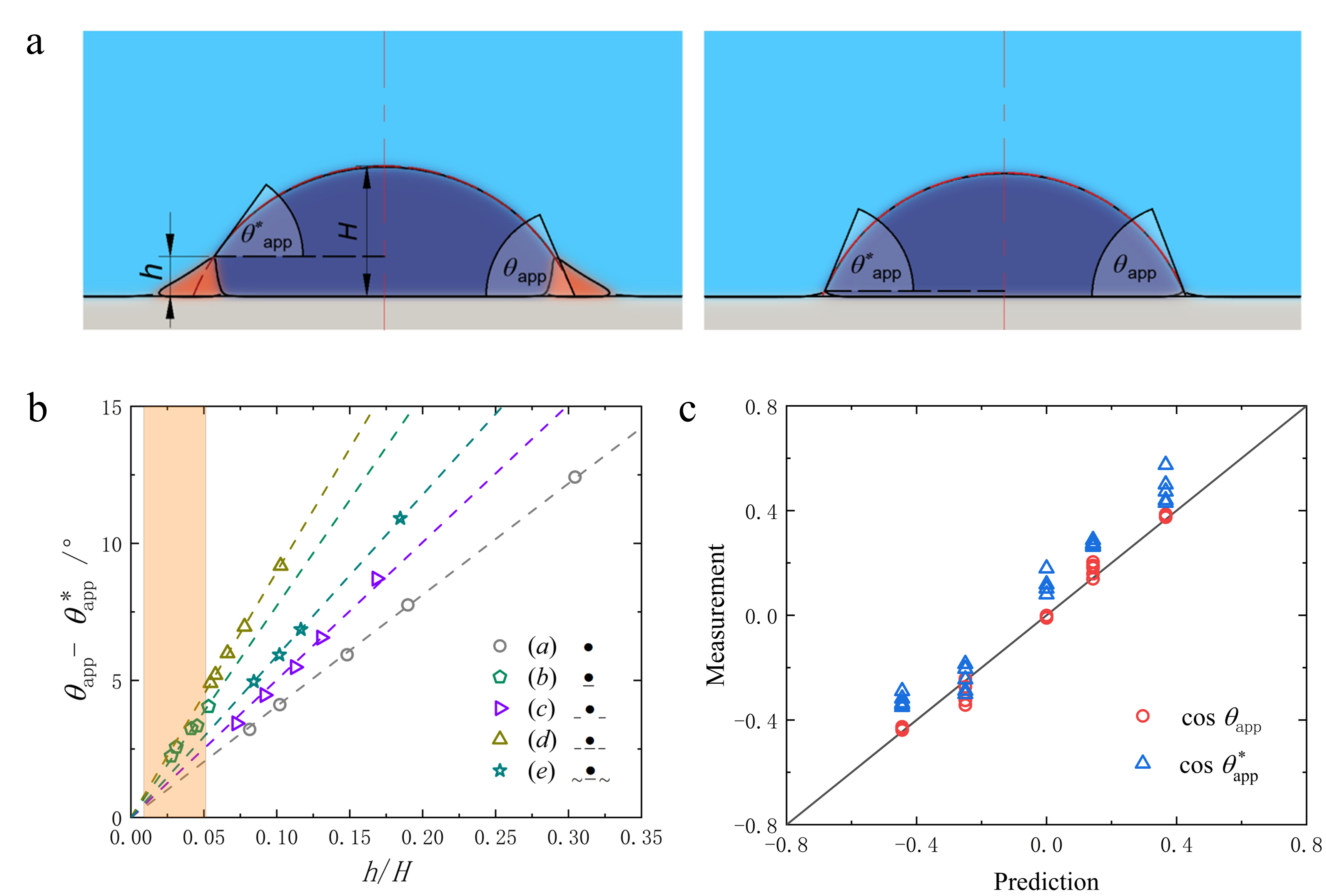}
    \captionsetup{justification=justified, singlelinecheck=false} 
    \caption{Variation of the apparent contact angle with lubricant ridge size. \textbf{a} Schematic illustrating the variation of $\theta_{\rm app}^*$ and $\theta_{\rm app}$ with the lubricant ridge size. \textbf{b} Difference between $\theta_{\rm app}$ and $\theta_{\rm app}^*$ as a function of $h/H$. The highlighted zone represents the range of $h/H$ observed in the experiments. \textbf{c} Comparison of $\theta_{\rm app}^*$ and $\theta_{\rm app}$ from \textbf{b} with theoretical predictions.}
    \label{fig:measurements of apparent contact angle}
\end{figure}

We carried out tests to characterize the effect of the size of the lubricant ridge on the apparent contact angle of the droplet. 
For a given patterned liquid surface, we varied the pressure constraint as a means to vary the size of the lubricant ridge.
We measured the apparent contact angles $\theta_{\rm app}$ and $\theta_{\rm app}^*$ using the definitions given in section IC.
As shown in Figure~\ref{fig:measurements of apparent contact angle}a, $\theta_{\rm app}^*$ increases with decreasing ridge size, $h$, defined as the height of the intersection of the ridge with the droplet surface. 
On the other hand, $\theta_{\rm app}$ does not vary significantly with ridge size. 
Figure~\ref{fig:measurements of apparent contact angle}b shows the difference between $\theta_{\rm app}$ and $\theta_{\rm app}^*$ vs $h/H$, where $H$ is the height of the droplet. We found that $\theta_{\rm app}^*$ approaches $\theta_{\rm app}$ and the fitted lines converge to zero as $h/H$ decreases. The highlighted zone represents the range of ridge height observed in our experiments.
%
%
Figure~\ref{fig:measurements of apparent contact angle}c shows the comparison of these simulation data with the theoretical prediction.
In all the cases shown here, the simulation data for $\theta_{\rm app}$ are in excellent agreement with the theoretical prediction.

